\documentclass{article}

\usepackage{amsmath,amsthm,amssymb,amsfonts}
\usepackage{booktabs}
\usepackage{algorithm}
\usepackage{algorithmic}
\usepackage{enumitem}
\usepackage[utf8]{inputenc}
\usepackage{hyperref}
\usepackage{graphicx}
\usepackage[numbers]{natbib}

\begin{document}

\title{COREclust: a new package for a robust and scalable analysis of complex data}

\author{Camille Champion$^{1}$, Anne-Claire Brunet$^{1}$\\ Jean-Michel Loubes$^{1}$, Laurent Risser$^{2}$}

\date{
  $\null^{1}$ Institut de Math\'ematiques de Toulouse (UMR 5219), Universit\'e de Toulouse, F-31062 Toulouse, France\\
  $\null^{2}$ Institut de Math\'ematiques de Toulouse (UMR 5219), CNRS, France\\
}
\maketitle

\begin{abstract}
In this paper, we present a new \textit{R} package \textit{COREclust} dedicated to the detection of representative variables in high dimensional spaces with a potentially limited number of observations. Variable sets detection is based on an original graph clustering strategy denoted CORE-clustering algorithm that 
 detects CORE-clusters, \textit{i.e.} variable sets having a user defined size range and in which each variable is very similar to at least another variable.
Representative variables are then robustely estimate as the CORE-cluster centers. 
This strategy is entirely coded in  \textit{C++} and wrapped by \textit{R} using the \textit{Rcpp} package. 
A particular effort has been dedicated to keep its algorithmic cost reasonable so that it can be used on large datasets.  
After motivating our work, we will explain the CORE-clustering algorithm as well as a greedy extension of this algorithm. We will then present how to use it and results obtained on synthetic and real data.
\end{abstract}

\section[Introduction]{Introduction}\label{sec:intro}

Discovering representative information in high dimensional spaces with a limited number of observations is a recurrent problem in data analysis. Heterogeneity between the variables behaviour and multiple similarities  between variable subsets make the analysis of complex systems an unobvious task. In this context, we developped a \textit{R} package that wraps \textit{C++} functions dedicated to the detection of CORE-clusters, \textit{i.e.} of variable sets having a controled size range and with similar observed behaviours. As developped hereafter, these CORE-clusters are of particular interest in complex systems as they represent different types of behaviours that can be robustely detected in the observations.

Complex systems are typically modeled as undirected weighted graphs (\cite{complex_structure}, \cite{networks}) where the nodes (vertices)  represent the variables and the edge weights are a measure of similarity between the variables. They cover a wide range of applications in various fields: social science \cite{Social}, technology \cite{Technological} or biology \cite{Biological}. Cluster structures in such complex networks are non-trivial topological features that are estimated using graph clustering methods. 
The choice of a specific graph clustering algorithm is often application dependent on the nature (structure, size) of the data. 

A wide variety of traditional methods were developed to treat the problem of graph clustering,
 which include graph partitioning, hierarchical clustering,  partitional clustering and spectral clustering 
\cite{Community_detect2}, \cite{Algo_comm}, \cite{Evaluation}, \cite{Graph_clust}, \cite{Spectral}, \cite{hierarchical}, \cite{kmeans} .
The first technique \cite{graph_partition} aims to find partitions such that inter\-partition edges are minimized and is based on strong assumptions (need to know the number of groups and their size). 
The second standard category of methods called hierarchical clustering \cite{hierarchical}  can be divided into two types of algorithms: agglomerative algorithms (start with a cluster by node, then iteratively merge the clusters) and divisive algorithms (start with a single cluster, then iteratively split it). Such techniques make little assumptions on the number and the size of the groups. However, they do not allow to select the best  graph partition. Moreover, based on a pairwise similarity matrix, their computational cost increases quickly with the number of nodes. Partitional clustering methods  include k-means (\cite{kmeans}), k-median strategy,  and enable to form a fixed number of groups based on a distance function. Lately, the last approach \cite{Spectral}  that can have highly accurate results on different data types, imposes a high cost of computing results. Those algorithms are mostly based on the eigen-decomposition of Laplacian matrices of either weighted or unweighted graphs. 

A common issue when using those methods on complex system is that each node/variable will be set in a cluster. For instance, a large amount of clusters will be computed if the clustering granularity is high. Some clusters will then have a high similarity rate between their nodes/observations and other ones will contain no pertinent information. Quite often, interesting relations will also be split into several clusters, making it tricky to identify a sparse number of pertinent variables. Conversely, a low granularity will lead to few large clusters with a high internal heterogeneous behaviours making it even harder to identify representative variables. 
In the important context where the  number of observations $n$ is lower that the observations dimension $p$, an even more critical issue is that traditional graph clustering techniques suffer from this high dimensionality manifested through an instability due to high complexity. 
Selected variable clusters will then be likely to be meaningless.
A regularization strategy is therefore mandatory and requires the detection of pertinent clusters of representative variables. 
 
In our approach, we gather a controled number of variables which present similarities.  
Those tightly groups, called cores, constitute highly connected subnetworks and do not necessarily cover all nodes/variables. Those cores all support a central variable position charcterized by the lowest average distance between the other variables in the same core.
Note that the notion of core in graph clustering is already used in k-means clustering algorithms \cite{Network}. 
As we will develop in section \ref{sec:methodo}, a major difference between k-means and our proposed approach is that we do not need to tune the total number of clusters but their minimal size, so only similar variables will be gathered in CORE-clusters.
Other graph clustering algorithms, based on k-core decomposition, also partition the graph into a maximal group of entities which are connected to at least k other entities in the group, were also developed:  In \cite{coreness2}, the  detection of such a partition involves repeatedly deleting all the nodes whose degree is less than k until no more such nodes exist. In particular, based on the concept of degeneracy of a network meaning that the largest value of k for which a k-core exists, it aims to decrease the computational time of high computational complexity graph clustering methods $(\mathcal{O}(p^3))$, where $p$ is the number of nodes/variables without any significant loss of accuracy, applying them on main subgraphs of the network. The algorithm takes as input a clustering coefficient that indicates the degree to which nodes tend to cluster together. The method of \cite{coreness}, that is implemented in the package igraph, is also related to the notion of coreness as it calculates hierarchically the core number for each node with a complexity in the order of $\mathcal{O}(p)$. A node's coreness is k if it belongs to a k-core but not (k+1)-core. 
Unlike those articles, CORE-clustering algorithm seeks to directly control the core size  from an optimized parameter (minimal core size) and not the number of cores to be detected.

Our strategy then requires the tuning of a single and intuitive parameter: the minimum number of nodes/variables in each core. Thus, the heart of our approach is a new graph clustering algorithm, called CORE Clustering, that robustly identifies groups of representative variables (CORE-clusters) of a high dimensional system with potentially few observations having similarities. 
Only highly connected nodes are clustered together according to a limited level of granularity. The algorithmic cost of our strategy is additionnaly limited and scales well to large datasets. This point of view has been originally considered in \cite{brevet}  and we present here an improved different version of this initial idea. The package \textit{COREclust} is implemented in \textit{C++}  and is interfaced with \textit{R} in order to make it possible to select clusters of leading players within a potentally large complex system. Our methodology is described in Section~\ref{sec:methodo}. More details about the package and its use are then given in Section~\ref{sec:package}.
Section~\ref{sec:results} finally presents results obtained using our package and concludes this paper.

\section{Statistical Methodology}\label{sec:methodo}

\subsection{Interactions model}

We consider the set of $n$ observations $I=\{X_1,\cdots,X_n\}$ of the studied system. Each observation is described by $p$ quantitative variables, so $X_{i}=(x_{i}^{1},\cdots,x_{i}^{p})$ for $i \in {1,\cdots,n}$. 
The representative variables in the system will be identified based on a quantitative description of the variable interactions. This requires to define a notion of distance between the observed variables. In this paper, we use standard Pearson correlations but other distance measures could be alternatively used. The interactions are then modelled as an undirected weigted graph where the nodes represent the variables and the edges represent the variable pairs having a non-negligible similarity score. Each edge has also a weight equals to the similarity score between the nodes/variables it connects. 

We denote $G(N,E)$ the undirected weigthed graph modeling the observed system. The set $N$ contains  $p$ nodes, each of them representing an observed variable, and the set  $E$ contains the weighted edges that model the similarity level of the  variable pairs. 

\subsection{CORE-clusters}

As mentioned section \ref{sec:intro}, the representative variables of the studied system are indirectely detected as the centers of CORE-clusters. 
In our work, this central variable is simply the one that has the lowest average distance to all other variables in the CORE-cluster. 
The central part of our strategy is then the detection of CORE-clusters of variables based on the graph $G(N,E)$. CORE-clusters are  groups of nodes/variables that are densely connected to each other. 
The notion of $k$-core introduced firstly by Seidman ~\cite{Network} for unweighted graphs in 1983, formally describe those cohesive groups.
To this end, let us consider the degree $deg(n)$ of a node $n \in N$ as the number of edges $e\in E$ incident  to itself. 
\paragraph{Definition 1} A subgraph H  of a graph G(N,E) is a $k$-core or a core of degree k if and only if H is a maximal subgraph such that $\delta(G)>=k$ where 
$\delta(G)=\min\limits_{n \in N} deg(n)$ is the minimum degree of $G(N,E)$ (we say that k is the rank of such a core).

One major limitation of the graph decomposition into cores is its design to work on unweigthed graphs. However, real networks are quite frequently weighted, and their weights describe important properties of the underlying systems. In such networks, nodes have (at least) two properties that can characterize them, their degree and their weight. In our approach, we will consider links with major weights.

\subsection{CORE-clusters detection}

In this paper, we developed the CORE-clustering algorithm based on the notion of CORE-clusters. In order to prevent instability in the process of core detection from one run to the other, the user has to tune the coefficient $\tau$ so that the size of the CORE-clusters to be detected will be in $\left\{\tau,2\tau-1\right\}$. Thus, the $k$-cores to be detected directly depends on this coefficient. The latter controls the level of regularisation when estimating the representative variables: Large values of  $\tau$ leads to the detections of large sets of strongly related variables.  In that case, the representative variables of these sets is likely to be meaningful even if $n<p$. However each CORE-cluster may also contain several variables that would have ideally been representative.
On the contrary, too small values of $\tau$ will be likely to detect all meaningful representative variables but also a non-negligible number of false positive  representative variables, especially if the observations are noisy or  if $n<p$.
Note that another optional parameter is the threshold on the minimal similarity observed between two variable to define an edge in $G(N,E)$. Using it with values stricly higher than $0$  allows to reduce the graph size.

An important characteristic of our algorithm is that it first strongly simplifies $G(N,E)$ using a maximum spanning tree strategy. The resulting graph $G(N,T)$ has the same nodes as $G(N,E)$  and the tree structure $T$ instead of the initial edges $E$, where $T$ is a subset of $E$. All node pairs are then still connected to each other but there is only a single path between each node pair. This strongly sparsifies the graph and makes possible the use of fast and scalable algorithms as described section~\ref{sec:compmethodo}. CORE-clusters detection strategy, which estimates strongly interconnected variable sets also makes sense on $G(N,T)$ as computing a maximum spanning tree aims to find a tree sturcture out of the edges $E$ with the highest possible weights in average.

\section{Computational Methodology}\label{sec:compmethodo}

\subsection{Main interactions estimation}\label{ssec:InteracEstim}

The very first step of our strategy is to quantify the similarity between the different observed variables. The similarity is first computed and represented as a dense graph  $G(N,E)$, where $N$ contains  $p$ nodes, each of them representing an observed variable, and  $E$ contains $K_E=p(p-1)/2$ undirected weighted edges that model a similarity level between all pairs of variables.
Various association measures can be used to quantify the similarity between variables $i$ and $j$, $1\leq i,j\leq p$. The most popular ones are Spearman or Pearson correlation measures, different notions of distances, or smooth correlation coefficients. We used the absolute value of Pearson's correlation coefficient in our package, so the weight between the nodes $i$ and $j$ is the Pearson's correlation coefficient between the absolute values of the observed variables $i$ and $j$. In this approach, we consider the variables with no connection in between which will have a correlation coefficient set to zero. 
The algorithmic cost of this estimation using a naive technique is $\mathcal{O}(n^2p)$ but it can be easily parallelized using divide and conquer algorithms for reasonably large datasets, as in \cite{RandallPhdThesis}. For large to very large datasets, correlations should be computed on sparse matrices, using \textit{e.g.} \cite{corsparseURL2018} to make this task computationally tractable. 

Once  $G(N,E)$ is computed, its complexity is strongly reduced to make  possible the efficient use of scalable clustering strategies on our data.

The maximum spanning tree of $G(N,E)$ is built by using Kruskal's algorithm \cite{KruskalPNAS56}. We denote $G(N,T)$ the resulting graph, where $T$ has a tree-like structure. Details of the algorithm are given in Alg.~\ref{alg:MST}. The algorithmic cost of the sort procedure (line 1) is $\mathcal{O}(K_E\log{(K_E)})$. 
Then, the for loop (lines 4 to 10) only scans one time the edges. In this for loop, the most demanding procedure is the propagation of label $L(N_i)$, line 8. Fortunately, the nodes on which the labels are propagated are connex to $N_j$ in $G(N,E)$. We can then use a Depth-First Search algorithm \cite{Tarjan72depthfirst} for that task, making the average performance of the for loop $\mathcal{O}(K_E\log{(p)})$.

\begin{algorithm}
\caption{Maximum spanning tree algorithm}
\label{alg:MST}
\begin{algorithmic}[1]
\REQUIRE Graph $G(N,E)$ with nodes $N_i$, $i \in {1,\cdots, p}$ and edges $E_k$, $k \in {1,\cdots, K_E}$.
\REQUIRE Weight of edge $E_k$ is $W(E_k)$.
\STATE Sort the edges by decreasing weights, so $W(E_1) \geq W(E_2)  \geq \cdots  \geq W(E_{K_E})$. 
\STATE Assign label $L(N_i)=i$ to each node $N(i)$.
\STATE Initiate an edge list $T$ as void.
\FOR{$k=1:K_E$}
\STATE We denote $N_i$ and $N_j$ the nodes linked by edge $E_k$.
\IF{$L(N_i)!=L(N_j)$}
\STATE Add edge  $E_k$ to the list $T$
\STATE Propagate the label $L(N_i)$ to the nodes that have label $L(N_j)$.
\ENDIF
\ENDFOR
\RETURN Graph with a tree structure $G(N,T)$.
\end{algorithmic}
\end{algorithm}

\subsection{CORE-clusters detection}

\begin{algorithm}
\caption{CORE-clustering algorithm}
\label{alg:CCA}
\begin{algorithmic}[1]
  \REQUIRE Graph $G(N,T)$ with nodes $N_i$, $i \in {1,\cdots, p}$ and edges $T_k$, $k \in {1,\cdots, K_T}$.
  \REQUIRE Weight of edge $T_k$ is $W(T_k)$.
  \REQUIRE Granularity coefficient $\tau$.
  \STATE \COMMENT{Initiate the algoritm}
  \STATE Sort the edges by increasing weights. 
  \STATE Assign label $L(N_i)=i$ to each node $N(i)$.
  \STATE Set $CORElabel=-1$.
  \STATE \COMMENT{CORE-clusters detection}
  \FOR{$k=1:K_T$}
    \STATE We denote $N_i$ and $N_j$ the nodes linked by edge $E_k$.
    \IF{$L(N_i)!=L(N_j)$}
      \STATE Propagate the label $L(N_i)$ to the nodes that have label $L(N_j)$.
      \IF{number of nodes with label $L(N_i) \in \{\tau,\cdots, 2\tau-1\}$}
        \STATE Label $CORElabel$ is given to the nodes with label $L(N_i)$ 
        \STATE $CORElabel=CORElabel-1$
      \ENDIF
    \ENDIF
  \ENDFOR
  \STATE \COMMENT{Post-treatment of the labels}
  \FOR{$n=1:N$}
      \STATE If $L(N_n)>0$ then $L(N_n)=0$ else $L(N_n)=-L(N_n)$
  \ENDFOR
  \RETURN Labels $L$.
\end{algorithmic}
\end{algorithm}

In contrast with the other clustering techniques, our CORE-clustering approach detects clusters having an explicitely controled granularity level, and only gathers nodes/variables with a behaviour considered as similar (see section \ref{ssec:InteracEstim}). 
Core structures are detected from the maximal spanning tree $G(N,T)$ by gathering iteratively its nodes $N$ in an order that depends on the edge weights in $T$. CORE-structures are identified when a group of gathered nodes has a size in $\{\tau,\cdots, 2\tau-1\}$, where $\tau$ is the parameter that controls the granularity level. Algorithm details are given Alg.~\ref{alg:CCA}.

To understand how Alg.~\ref{alg:CCA} detects clusters of coherent nodes, we emphasize that the CORE-clustering algorithm is applied on the maximum spanning tree $G(N,T)$ and not on $G(N,E)$. Only the most pertinent relations of each node $N_n$ are then taken into account, so the nodes with no pertinent relation will be randomly linked to each other in $T$. As Alg.~\ref{alg:CCA} will first gather these nodes, they will generate many small and meaningless clusters. Parameter $\tau$ should then not be too small to avoid considering these clusters as CORE-clusters. Then, the first pertinent clusters will be established using the edges of $T$ with larger weights, leading to pertinent node groups. Finally, the largest weights of $T$ are treated in the end of Alg.~\ref{alg:CCA} in order to split into several CORE-clusters the nodes related to the most influent nodes/variables. 

Remark that the algorithmic structures of Alg.~\ref{alg:MST} and Alg.~\ref{alg:CCA} are similar. However Alg.~\ref{alg:CCA} runs on $G(N,T)$ and not on $G(N,E)$. 
The number of edges $K_T$ in $G(N,T)$ is much lower than $K_E$ since $T$ has a tree structure and not a complete graph structure. It should indeed be slightly higher than $p$ \cite{SteeleMCS2002}, which is much lower than $K_E=p(p-1)/2$. Moreover the propagation algorithm (lines 9 and 11) will never propagate then labels on more than $2\tau-1$ nodes.  
The algorithmic cost of the sort procedure (line 1) is then $\mathcal{O}(K_T\log{(K_T)})$ and the average performance of the for loop $\mathcal{O}(K_T\log{(\tau)})$.

\subsection{A greedy alternative for CORE-clusters detection}

We propose an alternative strategy for the CORE-clusters detection algorithms: The edge treatment queue may be ordered by following  decreasing edge weights instead of increasing edge weights. The nearest edges will then be first gathered making coherent CORE-clusters as in Alg.~\ref{alg:CCA}, although one CORE-cluster may contain several key players. To avoid gathering noisy information, the for loop on the edges (line 6 of Alg.~\ref{alg:CCA}) should also stop before meaningless edges are treated. This strategy has a key interest: It can strongly reduce the computational time dedicated to Algs.~\ref{alg:MST} and \ref{alg:CCA}. By doing so, Alg.~\ref{alg:MST} and modified Alg.~\ref{alg:CCA} are purely equivalent to Alg.~\ref{alg:gCCA}.

\begin{algorithm}
\caption{Greedy CORE-clustering algorithm}
\label{alg:gCCA}
\begin{algorithmic}[1]
\REQUIRE Graph $G(N,E)$ with nodes $N_i$, $i \in {1,\cdots, p}$ and edges $E_k$, $k \in {1,\cdots, K_E}$.
\REQUIRE Weight of edge $E_k$ is $W(E_k)$.
\REQUIRE Granularity coefficient $\tau$ and edge number to scan $\gamma \tau$.
  \STATE Sort the edges by decreasing weights. 
  \STATE Assign label $L(N_i)=i$ to each node $N(i)$.
  \STATE Set $CORElabel=-1$.
  \FOR{$k=1:floor(\gamma \tau)$}
    \STATE We denote $N_i$ and $N_j$ the nodes linked by edge $E_k$.
    \IF{$L(N_i)!=L(N_j)$}
      \STATE Propagate the label $L(N_i)$ to the nodes that have label $L(N_j)$.
      \IF{number of nodes with label $L(N_i) \in \{\tau,\cdots, 2\tau-1\}$}
        \STATE Label $CORElabel$ is given to the nodes with label $L(N_i)$ 
        \STATE $CORElabel=CORElabel-1$
      \ENDIF
    \ENDIF
  \ENDFOR
  \FOR{$n=1:N$}
      \STATE If $L(N_n)>0$ then $L(N_n)=0$ else $L(N_n)=-L(N_n)$
  \ENDFOR
\RETURN Labels $L$.
\end{algorithmic}
\end{algorithm}

Again, the structure of this algorithm is very similar to the structure of the Maximum Spanning Tree strategy section~\ref{ssec:InteracEstim}. The algorithmic cost of the sort procedure (line 1) is $\mathcal{O}(K_E\log{(K_E)})$. Then, the loop lines 4 to 13 scans $floor(\gamma \tau)$ edges where $\gamma$ should be sufficiently high to capture the main CORE-clusters but also where $\gamma \tau << K_E$. Labels propagations in this loop (lines 7 and 9) is also limited to $2\tau-1$ nodes. The average performance of the for loop is therefore $\mathcal{O}(\gamma \tau \log{(\tau)})$.

\subsection{Central variables selection in CORE-clusters}

Once a CORE-cluster is identified, we use a straighforward strategy to select its central variable: the distance between all pairs of variables in each CORE-cluster is computed using a Dijkstra's algorithm \cite{Dijkistra}, \cite{Dijkistra2} in $G(N,E)$.
The central variable is then the one that has the lowest average distance to all other variables in the CORE-cluster. As the CORE-clusters have less than $2\tau$ nodes, the algorithmic cost of this procedure is $\mathcal{O}(\tau^2)$ times the number of detected CORE-clusters, which should remain low even for large datasets.

\section{Package description}\label{sec:package}

The package  \textit{COREclust} introduced in this paper can be freely downloaded at the url \url{https://sourceforge.net/projects/core-clustering/files/Rpackage_CoreClustering.tar} and is based on a \textit{R} script calling \textit{C++} functions.
It only requires preinstalling the Rcpp package \cite{EddelbuettelJSS2011} to automatically compile and call \textit{C++} functions. Note that the \textit{C++} algorithm only depends on standard libraries so it requires a standard \textit{C++} compiler and no specific library.
After its installation, the package can be loaded in \textit{R} by typing library("COREclust"). 

The latter contains a set of \textit{C++} and \textit{R} functions for detecting the main core structures within a large complex system as well as different tests for measuring the accuracy and the robustness of the algorithm developed for that task.

\subsection{C++ functions}\label{ssec:cppfcts}

The main \textit{C++} function of \textit{COREclust} called by \textit{R} is \textit{principalFunction}. Its first inputs are the name of the file containing the similarity level between the variables (\textit{arg\_InputFile}), the minimal size of the cores (\textit{arg\_MinCoreSize}) and the coefficient $\gamma$ (\textit{arg\_Gamma}). Note that  $\gamma$ is only used when performing the greedy algorithm. The next input is the name of the file containing the output variable labels (\textit{arg\_OutputFile}). The last one is finally a coefficient $1$ or $2$ that controls whether the standard or the greedy algorithm will be performed (\textit{greedyOrNot}). It is worth mentionning that the input \textit{csv} file is a correlation matrix generated by \textit{R} before executing the \textit{C++} code. Then the structure of \textit{principalFunction} is as described in Alg.~\ref{Alg:principalFunction}.

\begin{algorithm}
\caption{Overview of function \textit{principalFunction} in the \textit{C++} code.}
\label{Alg:principalFunction}
\begin{algorithmic}[1]
\REQUIRE \textit{InputFile}, \textit{OutputFile},  $\tau$ and  $\gamma$ (if greedy algorithm) 
  \STATE Instantiate two void graphs \textit{Graph} and \textit{MSP} of class \textit{CompactGraph}. 
  \IF{Standard algorithm}
      \STATE \textit{Graph.ReadInMatrix(InputFile)}: Read \textit{Graph} in a file \textit{InputFile} representing the connectivity matrix.
      \STATE \textit{MSP.MaximumSpanningTree(\&Graph)}: Initiate \textit{MSP} as the maximum spanning tree of \textit{Graph}.
      \STATE \textit{MSP.CoreClustering}($\tau$): Compute the CORE-clustering of \textit{Graph}.
      \STATE \textit{MSP.SaveLabels(OutputFile)}: Save the node labels in \textit{OutputFile}.
  \ENDIF
  \IF{Greedy alorithm}
      \STATE \textit{Graph.ReadInMatrix(InputFile)}: Read \textit{Graph} in a file \textit{InputFile} representing the connectivity matrix.
      \STATE \textit{Graph.CoreClustering\_Greedy}($\tau$,$\gamma\tau$): Compute the greedy CORE-clustering of \textit{Graph}.
      \STATE \textit{Graph.SaveLabels(OutputFile)}: Save the node labels in \textit{OutputFile}.
  \ENDIF
\end{algorithmic}
\end{algorithm}

Remark that the \textit{C++} code heavily depends on the class \textit{CompactGraph}, that is summarized in file \textit{CoreClust.h}. This class contains all variables to store undirected weighted graphs and node Labels. It also contains different methods (\textit{i.e.} class functions) to read undirected weighted graphs in a list of weighted edges or a squared matrix, to write a graph and its labels, as well as to perform the CORE-clustering strategies described in  Alg.~\ref{alg:CCA} and  Alg.~\ref{alg:gCCA}.
As mentioned section~\ref{sec:methodo} the scalability of our algorithms directly depend on the algorithm that sorts the graph edges depending of their weights.  To make this step computationally efficient, we have used in the \textit{SortByIncreasingWeight} and \textit{SortByDecreasingWeight} methods of \textit{CompactGraph} the reference \textit{C++} function \textit{std::sort}. For $N$ elements, this function performs $\mathcal{O}(N\log{(N)})$ comparisons in average until \textit{C++11}, and  $\mathcal{O}(N\log{(N)})$ comparisons after \textit{C++11}.

\subsection{R functions}

User interactions are made through a R script \textit{algo\_clust.R} that executes the \textit{C++} code of Alg.~\ref{Alg:principalFunction}. It first takes as inputs the name of a \textit{Rdata} or a \textit{csv} file (\textit{arg\_InputFile}). The remaining arguments are the minimal size of the cores (\textit{arg\_MinCoreSize}), the name of the file containing the output variable labels (\textit{arg\_OutputFile}) and optionally the $\gamma$ coefficient (\textit{arg\_Gamma}) that controls the edge number to be scanned to capture the main cores. If it is not called as argument, $\gamma$ is set to zero and  the fifth input argument of the \textit{principalFunction} is set to 1 (CORE-clustering algorithm will be applied on submit data). Otherwise, \textit{greedyOrNot} is set to 2.

Note that the script \textit{algo\_clust.R} contains a \textit{R} function \textit{corr\_Mat} that computes the similarity matrix of the observed data. It takes as arguments the name of the file containing the dataset \textit{Obs\_file\_name} and a coefficient \textit{FileType} equals to $1$ if \textit{Obs\_file\_name} is a \textit{csv} file and $2$ otherwise. In the first case, the absolute value of Pearson's correlation is computed using the function \textit{cor} of package \textit{stat} (\cite{statsURL2015}) and the associated adjacency matrix is built. In the second case, the input matrix that contains the observations is of type \textit{sparseMatrix} (out of the package \textit{Matrix} \cite{MatrixURL2017}). Then, the absolute value of Pearson's correlation is computed on its coefficients using the function \textit{corSparse} of package \textit{qlcMatrix} \cite{corsparseURL2018}. 

It also includes the function \textit{Rcpp::sourceCpp} from the package \textit{Rcpp} that parses the specified C++ file or source code. A shared library is then built and its exported functions and Rcpp modules are made available in the specified environment.

In the R file, the similarity matrix is first computed and saved in the \textit{csv} file \textit{corrMat.csv} that will be loaded by the  \textit{principalFunction} of the  \textit{C++} code.

\section{Results on synthetic data}\label{sec:results}

In order to study the ability of our algorithm to detect core structures and clusters of nodes from the graph, we simulated correlated data with clusters of varying size, density and level of noise, many different situations which may be found in any system. We also tested the accuracy and the robustness of our algorithm based on a specific similarity measure.

\subsection{Simulated networks}

Many complex networks are known to be scale free, meaning that the network has a little amount of highly connected nodes (hubs) and many poorly connected nodes. 
Simulating such networks can be seen as a process through which we first generate the profile of leader actors (hubs) and then the profile of the remaining ones around those hubs.

We consider here that the simulated individual profiles are divided in $K$ different clusters of size $n_{C_1}$,$n_{C_2}$,...,$n_{C_K}$. 
A simulated expression data set $X \in \mathbb{R}^{N\times p}$ is then composed of $p=n_{C1}+n_{C_2}+...+n_{C_K}$ actor profiles generated for the $K$ clusters and has the structure:
\begin{equation}
X=\left(\mathbf{x}^{(1)}_{C_1},...,\mathbf{x}^{(n_{C_1})}_{C_1},\mathbf{x}^{(1)}_{C_2},...,\mathbf{x}^{(n_{C_2})}_{C_2},...,\mathbf{x}^{(1)}_{C_K},...,\mathbf{x}^{(n_{C_K})}_{C_K}\right)
\end{equation}
Strategy for simulating actor profiles in a single cluster of size $n_C$ is:
\begin{enumerate}
  \item Generate the  profile $\mathbf{x}^{(1)} =(x_1^{(1)},x_2^{(1)},...,x_N^{(1)})^\prime$ 
 of one variable from a normal distribution, $x_i^{(1)}\sim \mathcal{N}(0,1),\forall i=1,2,...,N$. This profile defines the leader or hub profile.  
 \item Choose a minimum correlation $r_{min}$ and a maximum correlation $r_{max}$ between the leader actor and other actors in the cluster. 
 \item Generate $n_C-1$  profiles such that the correlation of the j-th profile $\mathbf{x}^{(j)}$ 
 with the  leader actor profile $\mathbf{x}^{(1)}$ is forall $j=2,3,...,n_C$ close to
 $$
 r_{j}=r_{min}+\Delta_r\left( 1-\frac{j}{n_C} \right),
 $$
 where $\Delta_r=r_{max}-r_{min}$.
 The $r_2$ correlation is close to $r_{max}$ and the $n_C$-th correlation is equal to $r_{min}+\Delta_r=r_{max}$.
 The j-th profile is also generated by adding a Gaussian noise term to the leader profile,
 to allow a correlation with the leader profile close to the required correlation $r_j$
 $$
  x_i^{(j)}= x_i^{(1)}+\sqrt{\left( \frac{1}{r_j^2}-1 \right)} \epsilon_i^{(j)},\forall i=1,2,...,N
 $$
 where $ \epsilon_i^{(j)}\sim\mathcal{N}(0,\alpha) $. 
\end{enumerate}
From the matrix including the profile of each actor of the system $S$, the coefficients $s_{ij}\in[0,1]$ of the $p\times p$ matrix of Pearson correlation will be computed and finally represent the edges $E$ of the simulated network $G(N,E)$.

\subsection{Performance evaluation on synthetic data}

Clustering results can be validated using two scores: 
The \textit{external validation} compares the clustering result with a ground truth result. 
\textit{Internal validation} also evaluates the clustering quality using the result itself plus the graph properties, and no ground truth data. It can therefore also be used on real-life data. The latter will be introduced in \ref{sec:application}.

\subsubsection{External evaluation of cores quality on synthetic data}

Let $X_i$, $i \in \{1,\cdots,p\}$ be the variables and $C_j$ ($j \in [1,K]$) be the ground-truth CORE-clusters of variables, typically on synthetic data. We also denote $\hat{C_k}$ ($k \in [1,\hat{K}]$) the CORE-clusters predicted by our algorithm. 
In order to evaluate the quality of the prediction, we compute a score $S$ defined as:
\begin{equation}\label{eq:extScore}
\forall j \in \left\lbrace1,\cdots,\hat{K} \right\rbrace, S_j=\displaystyle \max_{k \in [1,K]} Card( X^i \in C_k \cap \hat{C_j}) \,,
\end{equation}
where $ 1\leq i \leq p$ and $S=\frac{1}{p} \sum\limits_{j=1}^{\hat{K}} S_j$.

To compute this score, each $S_j$ is equal to $0$ if there is no overlap between $C_j$ and any $\hat{C_k}$, and is equal to the number of variables in $C_j$ if a $\hat{C_k}$ contains all the variables of $C_j$. We emphasize that a CORE-cluster contains between $\tau$ and $2\tau$ variables so, if the synthetic CORE-clusters have an appropriate size, no $\hat{C_k}$ should match with more than one $C_j$. A score $S$ equals to $1$ then means that a perfecly accurate estimation of the $C_j$ was reached, and the closer to 0 its values are the less accurate the CORE-clusters detection is. 

Note that other scores, such as the Gini index could have potentially been used. Contrary to standard external validation scores used in supervised clustering, we do not evaluate a clustering quality with a label given to all variables/nodes. Here we assess the detection of clusters, potentially on a subset of variables and do not want to penalize or favor the fact that some variables have no label. We then use the score Eq.~\eqref{eq:extScore}.

\subsection{Simulations}

\begin{figure}[h]
\begin{center}
\includegraphics[width=0.95\linewidth]{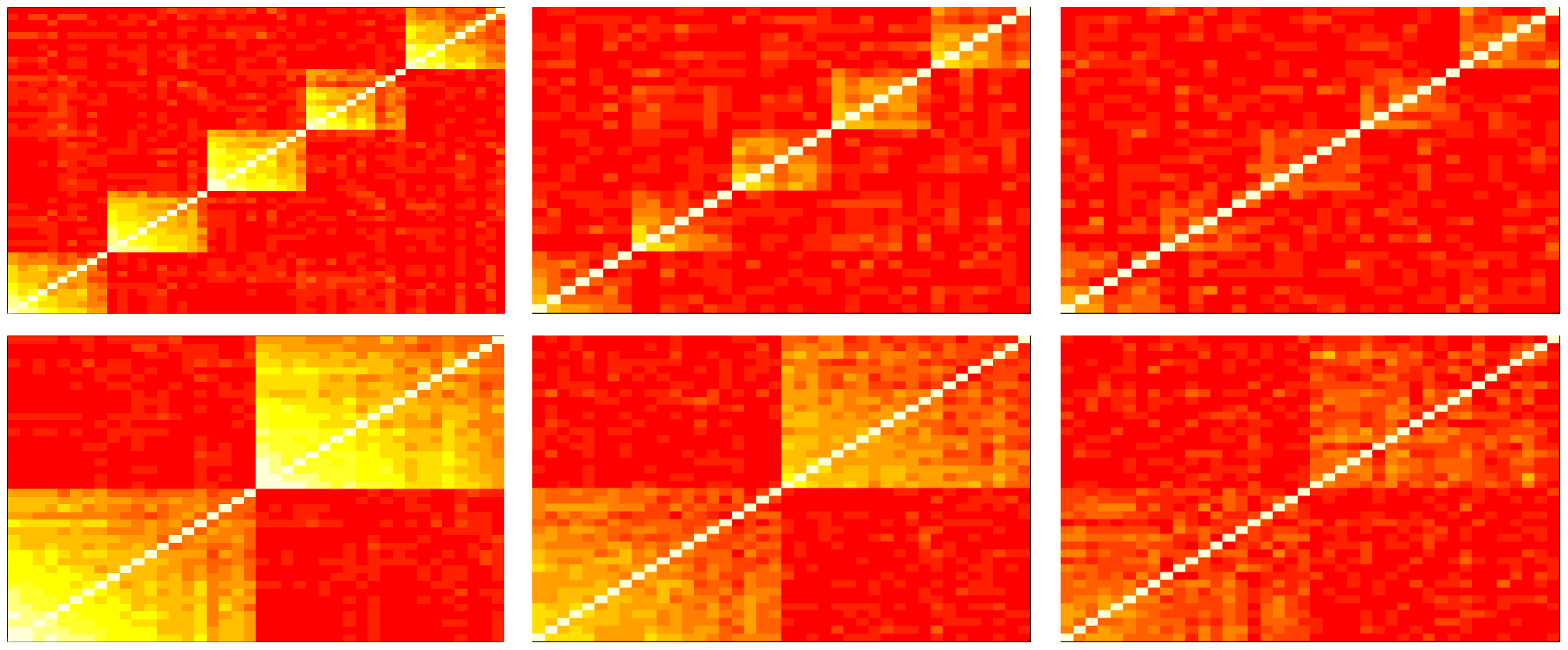}
\caption{Simulated co-expression matrices with  different noise levels and two or five clusters. \textbf{(From left to right)} Noise level $\alpha$ is 0, 3/4, 5/4. \textbf{(Top)} Two simulated clusters. \textbf{(Bottom)} Five simulated clusters.}
\label{simu}
\end{center}
\end{figure}

Standard and greedy algorithms were applied on four different simulations repeated $30$ times, to study their performances and robustness. On the basis of the built dataset  in section 4.1, each simulation was generated with well-defined clusters inside which correlations between actor profiles decrease linearly from $r_{max}=1$ to $r_{min}=0.5$.
Let simulation be the name of the function modeling the dataset introduced above.

\begin{enumerate}[label=\alph*)]
\item The first simulation matches the generation of the standard dataset of size $100\times p$ defined in section 4.1 with $K=2$  and $K=5$ clusters of size $(2,2)$ and $(7,\cdots,7)$; 
\item The second one consisted in creating noisy data and testing the CoreClustering algorithms on it. To this end, we simulated the standard dataset, and gradually added  to each variable an increasing gaussian noise term drawn from a normal distribution $\mathcal{N}(0,\alpha)$ depending on a coefficient $\alpha= \left\lbrace 0.25,1.5\right\rbrace$. Examples of simulated matrices with this noise term are shown in Fig.~\ref{simu};
\item The third simulation aims at changing the minimal size of core structures detected i.e the level of granularity by the algorithms along the process taking values from $(20,30,40,50)$ and checking their accuracy. The number of clusters was fixed at $K=5$ with sizes randomly drawn from the set $\left\lbrace50,60 \right\rbrace$;
\item The last simulation consists in varying  the size of the sample from 5 to 30 and setting $K=5$ and the minimal size of core structures to $20$.
\end{enumerate}

\subsection{Results}

\begin{figure}[h]
\begin{center}
\includegraphics[width=0.95\linewidth]{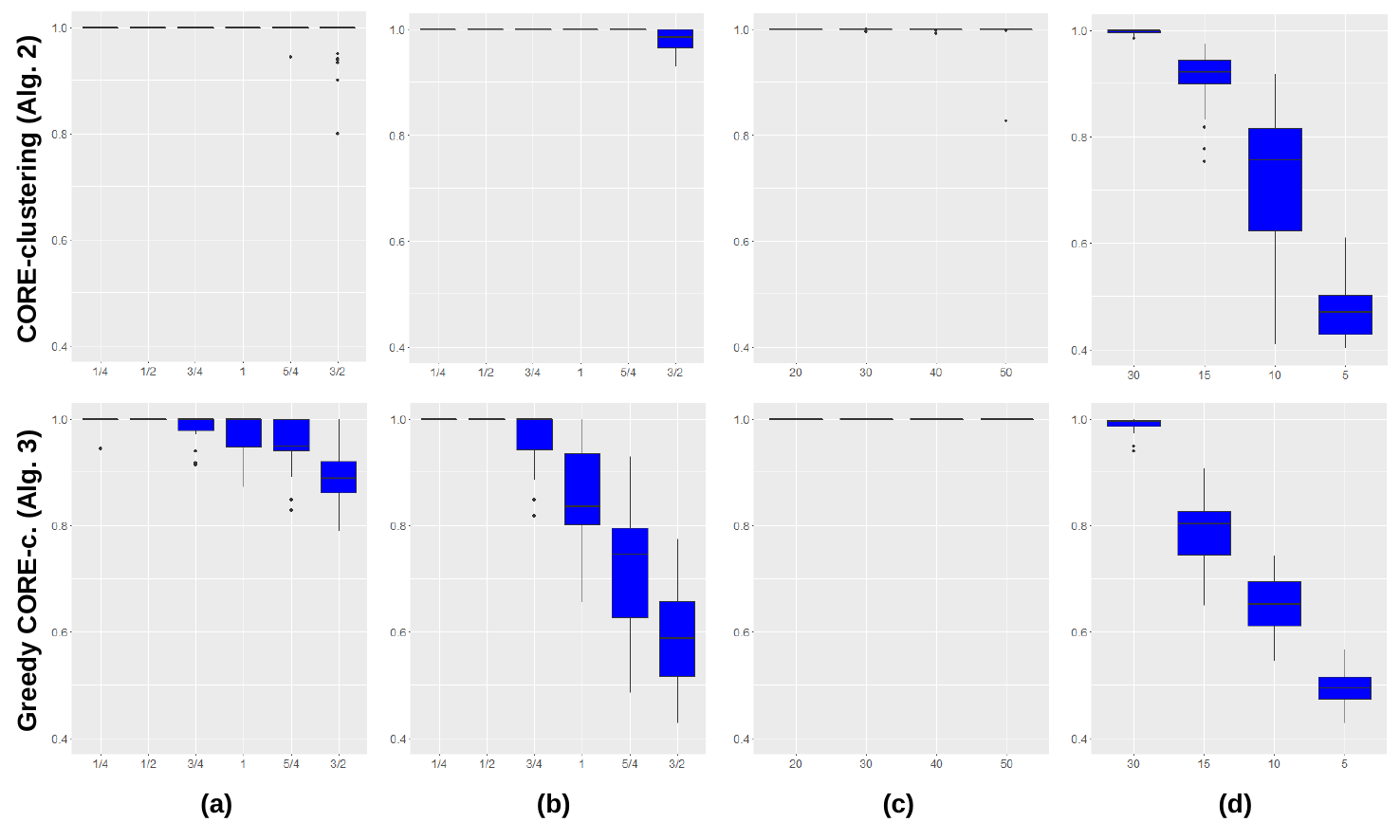}
\caption{Evaluation of the standard and the greedy CORE-clustering algorithm on simulated datasets.  \textbf{(a)} Two simulated clusters with noise levels $0.25$ to $1.5$; \textbf{(b)} Five simulated clusters with noise levels  $0.25$ to $1.5$; \textbf{(c)} Simulated datasets with five clusters of size randomly drawn form $50$ to $60$ and noise level $0.5$ with varying granularity coefficient $\tau$ from $20$ to $50$; \textbf{(d)} Five clusters simulated using 30, 15, 10 and 5 observations and noise level 3.}
\label{fig:results}
\end{center}
\end{figure}

For all the simulated datasets, CoreClustering and CoreClustering\_Greedy were applied on their matrices of correlation and then, the cores detected were compared to the real $K$ cluster groups thanks to the quality measurement. The results are shown in Fig.~\ref{fig:results} in the form of boxplots. The first four figures \textbf{(a)} and \textbf{(b)} show that the standard CoreClustering is slightely more robust to noisy dataset than the CoreClustering\_Greedy . The opposite effect can be observed when the sample size of the dataset decreases (\textbf{(d)}). Moreover, it should be noticed that increasing the number of clusters in the dataset slightly declines the accuracy of the detected cores. However, the two algorithms are really effective when varying the granularity coefficient (\textbf{(c)}) which differentiate it from other clustering techniques mentioned in the introduction. Nevertheless, it must be noted that the CoreClustering\_Greedy is really more effective in terms of processing time than the other one.

\section{Package application on real data}\label{sec:application}

The \textit{COREclust} package is distributed with example datasets, in particular a classic \cite{Yeast} dataset that can be found on the UCI Machine Learning Repository\footnote{\url{https://archive.ics.uci.edu/ml/datasets/Yeast}}. We present in this section results obtained using the \textit{COREclust} algorithms on this dataset.

\subsection{Data yeast}

Using simulated data, we showed that CORE-clustering algorithms performs well to detect core structures in a high dimensional network. This section illustrates the use of Core-Clustering algorithms on  the well-known synchronized yeast cell cycle data set of \cite{Yeast}, which includes 77 samples under various time during the cell cycle and a total of 6179 genes, of which 1660 genes are retained for this analysis after preprocessing and filtering. The goal of this analysis is then to detect core structures among the correlation patterns in the time series of gene expressions of yeast measured along the cell cycle. From this dataset introduced as a matrix, a measure of similarity between all gene pairs was measured with the absolute value of Pearson's correlation. A total of about $1.3\times10^6$ weighted edges will then be considered when representing the variable correlations in a graph.

\subsection{Sparse data yeast}

The second application of the two developed algorithms is based on a sparse version of the data yeast mentioned above. From the dataset built, $80\%$ of the information on it is set to zero, making the data sparse.  The matrix containing those data was then converted to a sparse matrix type using the package Matrix of R and a measure of similarity suitable for it was quantified using  \textit{e.g.} \cite{corsparseURL2018}.

\subsection{Command lines}

Those similarity matrices are then saved in a CSV format and passed as the first argument of the functions \textit{algo\textunderscore clust} and \textit{algo\textunderscore clust\textunderscore greedy}. The latter are finally run using the following command line \textit{Rscript algo\textunderscore clust.R in\textunderscore InputFile in\textunderscore MinCore\textunderscore Size in\textunderscore Outputfile} or \textit{Rscript algo\textunderscore clust\textunderscore greedy.R in\textunderscore InputFile in\textunderscore MinCore\textunderscore Size in\textunderscore Outputfile}, where the \textit{in\textunderscore MinCoreSize} and  \textit{in\textunderscore Outputfile} arguments are fixed by the user.

\subsection{Internal evaluation of the strength of the connections quality in cores}

We are interested here in evaluating the performances of the core detection algorithms using internal criteria. Let us intoduce the following evaluation metrics which will quantify the internal quality, i.e, the minimal property of the density of a core.

Let $W=(w_{ik})_{1\leq i,k\leq p}$ be the correlation matrix between each variable $X^1,...,X^p$. Let $\hat{K}$ the number of cores detected. Each variable $X^1,\cdots,X^p$ is assigned to a unique core $\hat{C_j}$ ($j \in [1,\hat{K}]$).
The quality measurement of internal connections strength in cores is defined as:
\begin{equation}\label{eq:intScore}
IC(\hat{K})=\frac{\displaystyle\min_{\left\lbrace i\in [1,p] | X_i\in \hat{C}_j\right\rbrace} \displaystyle\max_{\left\lbrace k\in [1,p] | X_k\in \hat{C}_j\right\rbrace} w_{i,k}}{\displaystyle\max_{1\leq i,k\leq p} w_{i,k}} \,,
\end{equation}
where we suppose that a weight equals to zero means that there is no connection between the two nodes involved. For each node of a given core, this score first determines the maximal strength of the edges that connect this node to its adjacent nodes. Over all those coefficents, the minimal  one is then divided by the maximal strength of all edges in the core. As  the last coefficient is very closed to one, the final score emphasizes the minimal strength of all vertices in each core between its internal nodes. 
A score $IC(\hat{K})$ closed to one then means that the algorithm reveals a core with highly connected nodes.

\subsection{Result}

\begin{center}
\begin{figure}[h]
\includegraphics[width=1.0\linewidth]{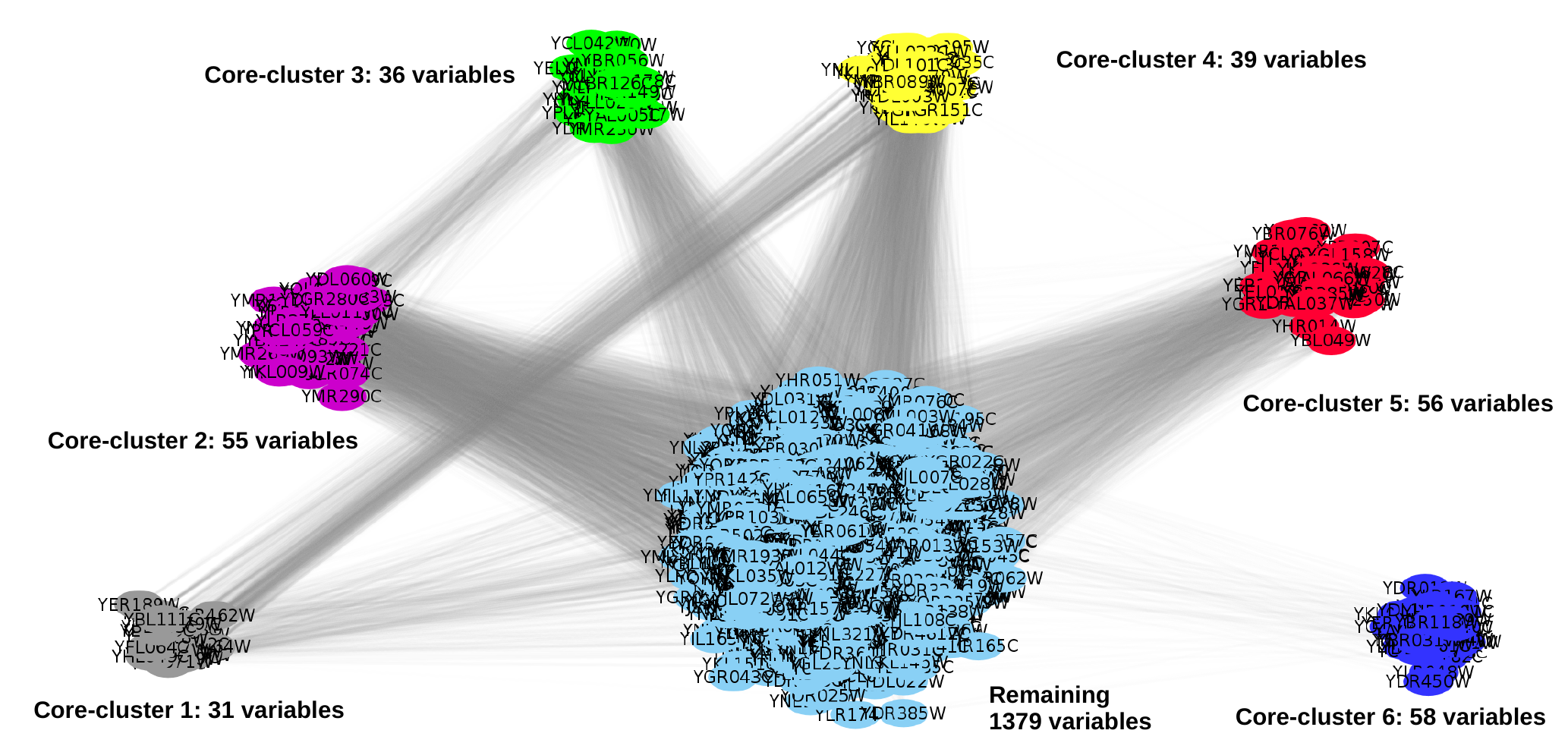}
\caption{CORE-clusters obtained using Alg.~2 on the yeast data}
\label{fig:network}
\end{figure}
\end{center}

In this context, the standard and greedy Core-Clustering algorithms were applied on the matrices of similarity introduced above with the aim of proving their performances on a real complex dataset. Applied on the original one, they detected respectively  $ 6$ and $25$ core structures  whereas $7$ and $37$ were catched when it comes to their sparse version. The minimal size of core structures was set at $30$. A visualization of the core clusters detected by the standard algorithm was made (Figure ~\ref{fig:network}) using Cytoscape\footnote{\url{http://www.cytoscape.org/}} where the transparency on the edges linearly depends on the correlation between the variables. Transparency for correlation 0.5 is $1\%$ and transparency for correlation 1 is $20\%$. The quality of the connections strength  in core structures detected in each dataset were then internally evaluated  using the quality measurement introduced in section \ref{sec:results}. The mean of the scores $IC(\hat{K})$ obtained for each cores when the standard algorithm is applied on the first type of data is  in $[0.70,0.84]$ whereas it is in $[0.64,0.87]$ for the greedy algorithm. Thus, the assumption that CORE-clusters are sets of highly connected variables is verified. For the sparse version of the data yeast, the scores  $IC(\hat{K})$ are respectively in $[0.69,0.73]$  and in  $[0.52,0.77]$. Those low scores could be dued to the limited number of information in the sparse data. Additionnaly, the core structures detected in each dataset were compared using the quality measurement of clusters introduced in Section \ref{sec:results}. The scoring was about $S_j=0.93$  for the first type of data and $S_j=0.95$ for its sparse version. However, in view of the different number of core structures detected by each algorithm, this measurement actually compares the first cores recognized by its greedy version with all of the standard one. Thus, the resultant scores show that the two algorithms are both robust when they are applied on a complex dataset and performs efficiently even on very sparse dataset. The computational times for the standard  and greedy version of the clustering algorithm applied on the original dataset were respectively about $319.60$ and $26.86$ seconds on a Core i5 computer with a RAM of $4.00$Go. Then, when they were tested on a sparse dataset, the computational times were about 255.92 and 31.78 seconds.

\section{Conclusion}

Although complex systems in high dimensional spaces with a limited number of observations are quite common across many fields, efficient methods to treat the associated problem of graph clustering is an unobvious task. Some of these techniques, based on assumptions in view of controlling the variables contribution to the global clustering, often do not allow to select the best graph partition. In reply to this issue, we developped the \textit{R} package \textit{COREclust} introduced in this paper. This package offers two strategies to robustly identify groups of representative variables of the studied system by tuning a single intuitive parameter: the minimum number of variables in each core. Its effectiveness  as a graph clustering technique was further  illustrated by  simulations and applications on real dataset.

\end{document}